\documentclass[conference,10pt]{IEEEtran}

\usepackage[dvips]{graphicx}
\usepackage[english]{babel}
\usepackage[latin1]{inputenc}
\usepackage[leqno]{amsmath}
\usepackage{amssymb}
\usepackage{subfig}
\usepackage{url}

\usepackage{cite} 

\newtheorem{theorem}{Theorem}
\newtheorem{lemma}{Lemma}

\newcounter{MYtempeqncnt}

\begin{document}

\title{One shot schemes for decentralized quickest change detection}
\author{
\authorblockN{Olympia Hadjiliadis}
\authorblockA{Department of Mathematics\\
Brooklyn College, C.U.N.Y. \\
Email: ohadjiliadis@brooklyn.cuny.edu} \\
\and
\authorblockN{Hongzhong Zhang}
\authorblockA{Department of Mathematics\\
Graduate Center, C.U.N.Y.\\
Email: hzhang3@gc.cuny.edu} \\
\and
\authorblockN{H. Vincent Poor }
\authorblockA{Department of Electrical Engineering\\
Princeton University\\
 Email: poor@princeton.edu}}

\maketitle

\selectlanguage{english}

\begin{minipage}{\textwidth}
\vspace*{-10ex}
\begin{center}
DISTRIBUTED INFERENCE AND DECISION-MAKING IN MULTISENSOR SYSTEMS,\\
ORGANIZERS: ALEXANDER TARTAKOVSKY AND VENUGOPAL VEERAVALLI.
\end{center}
\end{minipage}

\begin{abstract}
This work considers the problem of quickest detection with $N$ distributed sensors
that receive continuous sequential observations from the environment. These sensors
employ cumulative sum (CUSUM) strategies and communicate to a central fusion center
by one shot schemes. One shot schemes are schemes in which the sensors communicate
with the fusion center only once, after which they must signal a detection. The
communication is clearly asynchronous and the case is considered in which the fusion
center employs a minimal strategy, which means that it declares an alarm when the
first communication takes place. It is assumed that the observations received at the
sensors are independent and that the time points at which the appearance of a signal
can take place are different. It is shown that there is no loss of performance of
one shot schemes as compared to the centralized case in an extended Lorden min-max
sense, since the minimum of $N$ CUSUMs is asymptotically optimal as the mean time
between false alarms increases without bound.
\end{abstract}
\vspace*{2ex} \noindent {\bf Keywords: One shot schemes, CUSUM, quickest
detection\footnote{This research was supported in part by the U.S. National Science
Foundation under Grants ANI-03-38807, CNS-06-25637 and CCF-07-28208}}

%
\IEEEpeerreviewmaketitle
\vspace*{5ex}
\section{Introduction}
\label{sec:Introduction}
\begin{figure}
  \centering
  \includegraphics[width=.5\textwidth]{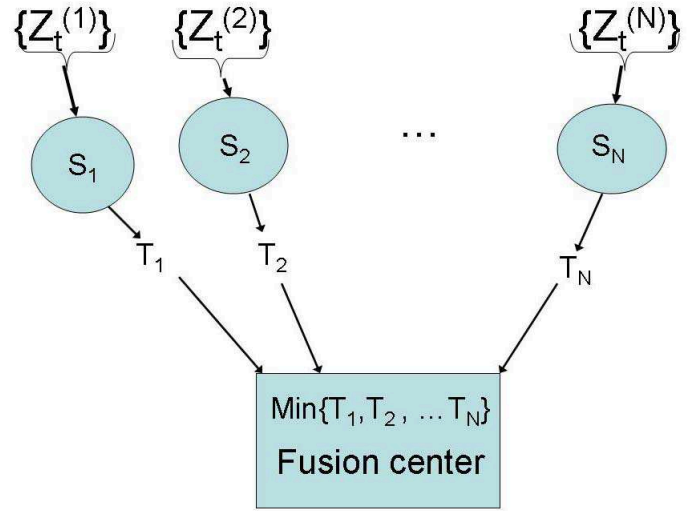}
  \caption{One shot communication in a decentralized system of $N$ sensors.}
  \label{fig1}
\end{figure}

The problem of decentralized sequential detection with data fusion dates back to the
1980s with the works of \cite{Al-IbVars88} and \cite{Al-IbVars89}. We are interested
in the problem of quickest detection in an $N$-sensor network in which the
information available is distributed and decentralized, a problem introduced in
\cite{Veer01}. We consider the situation in which the onset of a signal can occur at
different times in the $N$ sensors, that is the change points can be different for
each of the $N$ sensors. We assume that each sensor runs a cumulative sum (CUSUM)
algorithm as suggested in \cite{Mous06,TartKim06,TartVeer03,TartVeer04,TartVeer08}
and communicates with a central fusion center only when it is ready to signal an
alarm. In other words, each sensor communicates with the central fusion center
through a one shot scheme. We assume that the $N$ sensors receive independent
observations, which constitutes an assumption consistent with the fact that the $N$
change points can be different. So far in the literature (see
\cite{Mous06,TartKim06,TartVeer03,TartVeer04,TartVeer08}) it has been assumed that
the change points are the same across sensors. In this paper we consider the case in
which the central fusion center employs a minimal strategy, that is, it reacts when
the first communication from the sensors takes place. We demonstrate that, in the
situation described above, at least asymptotically, there is no loss of information
at the fusion center by employing the minimal one shot scheme. That is, we
demonstrate that the minimum of $N$ CUSUMs is asymptotically optimal in detecting
the minimum of the $N$ different change points, as the mean time between false
alarms tends to $\infty$, with respect to an appropriately extended Lorden criterion
\cite{Lord} that incorporates the possibility of $N$ different change points. As an
observation model we consider a continuous time Brownian motion model, which is a
good approximation to reality for measurements taken at a high rate. Moreover, given
a high rate of observations from any distribution, the central limit theorem asserts
that sums of such observations are normally distributed and therefore the Brownian
motion model is a plausible model for such situations.

The communication structure considered in this paper is summarized in Figure
\ref{fig1}, in which $T_i$  for $i=1,\ldots, N$ denote stopping times associated
with alarms at sensors $S_i$ $i=1,\ldots, N$, respectively.

In the next section we formulate the problem and demonstrate asymptotic optimality
(as the mean time between false alarms tends to $\infty$), in an extended min-max
Lorden sense, of the minimum of $N$ CUSUM stopping times in the case of centralized
detection. We then argue that this result suggests no loss in performance of the one
shot minimal strategy employed by the fusion center in the case of decentralized
detection. We finally discuss an extension of these results to the case of
correlated sensors.

\section{The centralized problem}
We sequentially  observe the processes $\{\xi_t^{(i)};t \ge 0\}$ for all
$i=1,\ldots,N$ with the following dynamics:
\begin{eqnarray}
\label{corr} d\xi_t^{(i)} = \left \{
\begin{array}{ll}
dw_t^{(i)}  & t \le \tau_i \\
\mu\,dt + dw_t^{(i)}&  t > \tau_i,
\end{array}
\right.
\end{eqnarray}

\noindent where $\mu>0$ is known \footnote{Due to the symmetry of Brownian motion,
without loss of generality, we can assume that $\mu>0$.}, $\{w_t^{(i)}\}$ are
independent standard Brownian motions, and the $\tau_i$'s are unknown constants.

An appropriate measurable space is $\Omega=C[0,\infty) \times C[0,\infty) \times
\ldots \times C[0,\infty)$ and $\mathcal{F}=\cup_{t
> 0} \mathcal{F}_t$, where $\{\mathcal{F}_t\}$ is the filtration of the observations
with $\mathcal{F}_t=\sigma\{s \le t; (\xi_s^{(1)},\ldots,\xi_s^{(N)})\}$. Notice
that in the case of centralized detection the filtration consists of the totality of
the observations that have been received up until the specific point in time $t$.

On this space, we have the following family of probability measures
$\{P_{\tau_1,\ldots,\tau_N}\}$, where $P_{\tau_1,\ldots,\tau_N}$ corresponds to the
measure generated on $\Omega$ by the processes $(\xi_t^{(1)},\ldots,\xi_t^{(N)})$
when the change in the $N$-tuple process occurs at time point $\tau_i$,
$i=1,\ldots,N$. Notice that the measure $P_{\infty,\ldots,\infty}$ corresponds to
the measure generated on $\Omega$ by $N$ independent Brownian motions.

Our objective is to find a stopping rule $T$ that balances the trade-off between a
small detection delay subject to a lower bound on the mean-time between false alarms
and will ultimately detect $\min\{\tau_1,\ldots,\tau_N\}$ \footnote{In what follows
we will use $\tau_1 \wedge \ldots \wedge \tau_N$ to denote
$\min\{\tau_1,\ldots,\tau_N\}$.}.

 As a performance measure we consider
\begin{multline}
\label{extLorden} J_M^{(N)}(T)=
\\\sup_{\tau_1,\ldots,\tau_N}\textrm{essup}\,E_{\tau_1,\ldots,\tau_N}
\left\{(T-\tau_1 \wedge \ldots \wedge \tau_N\})^{+} | \mathcal{F}_{\tau_1 \wedge
\ldots \wedge \tau_N}\right\}
\end{multline} where the supremum over $\tau_1, \ldots, \tau_N$ is taken over
the set in which $\min\{\tau_1,\ldots,\tau_N\} <\infty$. That is, we consider the
worst detection delay over all possible realizations of paths of the $N$-tuple of
stochastic processes $(\xi_t^{(1)},\ldots,\xi_t^{(N)})$ up to
$\min\{\tau_1,\ldots,\tau_N\}$ and then consider the worst detection delay over all
possible $N$-tuples $\{\tau_1,\ldots,\tau_N\}$ over a set in which at least one of
them is forced to take a finite value. This is because $T$ is a stopping rule meant
to detect the minimum of the $N$ change points and therefore if one of the $N$
processes undergoes a regime change, any unit of time by which $T$ delays in
reacting, should be counted towards the detection delay.

The criterion presented in (\ref{extLorden}) results in the corresponding stochastic
optimization problem of the form:
\begin{equation}
\begin{array}{c}
\displaystyle \inf_T J_M^{(N)}(T)
\\
\textrm{subject to}~~~ E_{\infty,\ldots,\infty} \left\{T \right\}  \geq   \gamma.
\end{array} \label{eqnproblem}
\end{equation}

We notice that the expectation in the above constraint is taken under the measure
$P_{\infty,\ldots,\infty}$. This is the measure generated on the space $\Omega$ in
the case that none of the $N$ processes $(\xi_t^{(1)},\ldots,\xi_t^{(N)})$ changes
regime. Therefore, $E_{\infty,\ldots,\infty} \left\{T \right\}$ is the mean time
between false alarms.

In the case of the presence of only one stochastic process (say $\{\xi_t^{(1)}\}$),
the problem becomes one of detecting a one-sided change in a sequence of Brownian
observations, or a vector of observations $(\xi_t^{(1)}, \ldots, \xi_t^{(N)})$ with
the same change points, whose optimal solution was found in \cite{Beib} and
\cite{Shir96}. The optimal solution is the continuous time version of Page's CUSUM
stopping rule, namely the first passage time of the process
\begin{eqnarray}
\label{CUSUMprocess} y_t^{(1)} & = &\sup_{0 \le \tau_1 \le t}\log
\left.\frac{dP_{\tau_1}}{dP_\infty}\right|_{\mathcal{F}_t}=u_t^{(1)}-m_t^{(1)},~\textrm{where} \\
\label{u} u_t^{(1)} & = & \mu \xi_t^{(1)} -\frac{1}{2} \mu^2 t,
\end{eqnarray} and
\begin{eqnarray}
\label{m} m_t^{(1)} & = & \inf_{0 \le s \le t} u_s^{(1)}.
\end{eqnarray}

\noindent The CUSUM stopping rule is thus
\begin{eqnarray}
\label{CUSUMstop} T_{\nu} & = & \inf\{t \ge 0; y_t^{(1)} \ge \nu\},
\end{eqnarray}

\noindent where $\nu$ is chosen so that $E_\infty\left\{T_{\nu}\right\}\equiv
\frac{2}{\mu^2}f(\nu)=\gamma$, with $f(\nu)=e^{\nu}-\nu-1$ (see for example
\cite{Hadj06}) and
\begin{eqnarray}
\label{DD} J_M^{(1)}(T_{\nu})\equiv E_0\left\{T_{\nu}\right\}=
\frac{2}{\mu^2}f(-\nu).
\end{eqnarray}
The fact that the worst detection delay is the same as that incurred in the case in
which the change point is exactly $0$ is a consequence of the non-negativity of the
CUSUM process, from which it follows that the worst detection delay occurs when the
CUSUM process at the time of the change is at $0$ \cite{Hadj06}.

We remark here that if the $N$ change points were the same then the problem
(\ref{eqnproblem}) is equivalent to observing only one stochastic process which is
now $N$-dimensional. Thus, in this case, the detection delay and mean time between
false alarms are given by the formulas in the above paragraph.

Returning to problem (\ref{eqnproblem}), it is easily seen that in seeking solutions
to this problem, we can restrict our attention to stopping times that achieve the
false alarm constraint with equality \cite{Mous86}. The optimality of the CUSUM
stopping rule in the presence of only one observation process suggests that a CUSUM
type of stopping rule might display similar optimality properties in the case of
multiple observation processes. In particular, an intuitively appealing rule, when
the detection of $\min\{\tau_1,\ldots,\tau_N\}$ is of interest, is $T_{h}=T_{h}^1
\wedge \ldots \wedge T_{h}^N$, where $T_h^i$ is the CUSUM stopping rule for the
process $\{\xi_t^{(i)};t \ge 0\}$ for $i=1,\ldots,N$. That is, we use what is known
as a multi-chart CUSUM stopping time \cite{Tart05}, which can be written as
\begin{eqnarray}
\label{CUSUMmultichart} T_{h} & = & \inf\left\{t \ge 0;
\max\{y_t^{(1)},\ldots,y_t^{(N)}\} \ge h\right\},
\end{eqnarray}

\noindent where $$y_t^{(i)}=\sup_{0 \le \tau_i \le t}\left.\log
\frac{dP_{\tau_i}}{dP_\infty}\right|_{\mathcal{F}_t}=\mu
\xi_t^{(i)}-\frac{1}{2}\mu^2 t-\inf_{s \le t} \left(\mu \xi_s^{(i)}-\frac{1}{2}
\mu^2 s\right).$$

\noindent and the $P_{\tau_i}$ are the restrictions of the measure
$P_{\tau_1,\ldots,\tau_N}$ to $C[0,\infty)$.

It is easily seen that
\begin{eqnarray*}
J_M^{(N)}(T_{h})=E_{0,\infty,\ldots,\infty}\left\{T_{h}\right\} & = &
E_{\infty,0,\infty,\ldots,\infty}\left\{T_h\right\}
\\ & = & \ldots \\& = & E_{\infty,\ldots,\infty,0}\left\{T_h\right\}.
\end{eqnarray*}

 This is because the worst detection delay occurs when at least one of the $N$
processes does not change regime. The reason for this lies in the fact that the
CUSUM process is a monotone function of $\mu$, resulting in a longer on average
passage time if $\mu=0$ \cite{HadjPoor}. That is, the worst detection delay will
occur when none of the other processes changes regime and due to the non-negativity
of the CUSUM process the worst detection delay will occur when the remaining one
processes is exactly at $0$.

Notice that the threshold $h$ is used for the multi-chart CUSUM stopping rule
(\ref{CUSUMmultichart}) in order to distinguish it from $\nu$ the threshold used for
the one sided CUSUM stopping rule (\ref{CUSUMstop}).

In what follows we will demonstrate asymptotic optimality of (\ref{CUSUMmultichart})
as $\gamma \to \infty$. In view of the discussion in the previous paragraph, in
order to assess the optimality properties of the multi-chart CUSUM rule
(\ref{CUSUMmultichart}) we will thus need to begin by evaluating
$E_{0,\infty,\ldots,\infty}\left\{T_h\right\}$ and
$E_{\infty,\ldots,\infty}\left\{T_h\right\}$.

Since the processes $\xi_t^{(i)}$,  $i=1,\ldots,N$, are independent it is possible
to obtain a closed form expression through the formula
\begin{eqnarray}
\label{E0INFTY}
\end{eqnarray}
\begin{multline}
  E_{0,\infty,\ldots,\infty}\left\{T_h\right\} = \int_0^\infty
P_{0,\infty,\ldots,\infty}(T_h>t) \\ \nonumber =  \int_0^\infty
P_{0,\infty,\ldots,\infty}(\{T_h^{1}>t\} \cap \ldots \cap \{T_h^N>t\}) dt
\\  =  \int_0^\infty P_0(T_h^1>t) \left[P_\infty(T_h^1>t)\right]^{N-1}dt.
\end{multline}

\noindent Similarly,
\begin{eqnarray}
\label{EINFTY}  E_{\infty,\ldots,\infty}\left\{T_h\right\} & = & \int_0^\infty
\left[P_\infty(T_h^1>t)\right]^N dt,
\end{eqnarray}

\noindent where $\{T_h^{i}>t\}=\{\sup_{0 \le s \le t}y_s^{(i)}
<h\}$. In other words, the evaluation of (\ref{E0INFTY}) and
(\ref{EINFTY}) is possible through the probability density function
of the random variable $\sup_{0 \le s \le t}y_s^{(i)}$ for arbitrary
fixed $t$ which appears in \cite{Malik}.

In order to demonstrate asymptotic optimality of (\ref{CUSUMmultichart}) we bound
the detection delay $J_M^{(N)}$ of the unknown optimal stopping rule $T^*$ by
\begin{eqnarray}\label{UB}
E_{0,\infty,\ldots,\infty}\left\{T_h \right\} & > & J_M^{(N)}(T^*),
\end{eqnarray}
where $h$ is chosen so that
\begin{eqnarray}\label{FACSh} E_{\infty,\ldots, \infty}\{T_h\}& = & \gamma.\end{eqnarray}
It is also obvious that $J_M^{(N)}(T^*)$ is bounded from below by the detection
delay of the one CUSUM when there is only one observation process, in view of the
fact that
\begin{eqnarray*}
& \sup_{\tau_1,\ldots,\tau_N}\textrm{essup}E_{\tau_1,\ldots,\tau_N}
\left\{{(T-\tau_1 \wedge \ldots \wedge \tau_N)}^{+} | \mathcal{F}_{\tau_1 \wedge
\ldots \wedge \tau_N} \right\} \ge &
\\ & \ge \sup_{\tau_1}\textrm{essup}\,E_{\tau_1} \left\{{(T-\tau_1)}^{+} |
\mathcal{F}_{\tau_1} \right\}. &
\end{eqnarray*}
The stopping time that minimizes $\sup_{\tau_1}\textrm{essup}\,E_{\tau_1}
\left\{{(T-\tau_1)}^{+} | \mathcal{F}_{\tau_1} \right\}$ is the CUSUM stopping rule
$T_\nu$ of (\ref{CUSUMstop}), with $\nu$ chosen so as to satisfy
\begin{eqnarray}\label{FACSnu} E_\infty\{T_\nu\}& = & \gamma.\end{eqnarray}
We will demonstrate that the difference between the
upper and lower bounds
\begin{eqnarray}
\label{ULB} E_{0,\infty,\ldots,\infty}\left\{T_h \right\} & >  J_M^{(N)}(T^*) & >
E_0\{T_\nu\},
\end{eqnarray}
is bounded by a constant as $\gamma \to \infty$, with $h$ and $\nu$
satisfying (\ref{FACSh}) and (\ref{FACSnu}), respectively.

\begin{lemma}
\label{asymptotics} We have
\begin{eqnarray}\label{AsymptoticSh}  \nonumber E_{0,\infty,\ldots,\infty}\left\{T_h \right\} & = & \frac{2}{\mu^2}\left[\log \gamma +\log
\frac{N\mu^2}{2}-1 + o(1)\right] ,\\&~&
\end{eqnarray} as $\gamma \to \infty$,
\end{lemma}
{\bf Proof:} Please refer to the Appendix for a sketch of  the proof.

Moreover, it is easily seen from (\ref{DD}) that
\begin{eqnarray}\label{AsymptoticSn} E_0\left\{T_\nu\right\} & = & \frac{2}{\mu^2}\left[\log
\gamma +\log \frac{\mu^2}{2} -1+o(1)\right].
\end{eqnarray}
Thus we have the following result.
\begin{theorem}
\label{main} The difference in detection delay $J_M^{(N)}$ of the unknown optimal
stopping rule $T^*$ and the detection delay of $T_h$ of (\ref{CUSUMmultichart}) with
$h$ satisfying (\ref{FACSh}) is bounded above by $$\frac{2}{\mu^2}\left(\log
N\right),$$ as $\gamma\to \infty$.
\end{theorem}
{\bf Proof:} The proof follows from Lemma \ref{asymptotics} and
(\ref{AsymptoticSn}).

\begin{figure*}[!t]
\begin{minipage}{\textwidth}
\begin{center}
{\bf\large The upper and lower bounds on the detection delay (DD) for the optimal
stopping rule}
\end{center}
\end{minipage}
\centerline{\subfloat[$\mu=\frac{1}{2}$]{\includegraphics[width =2.5in]{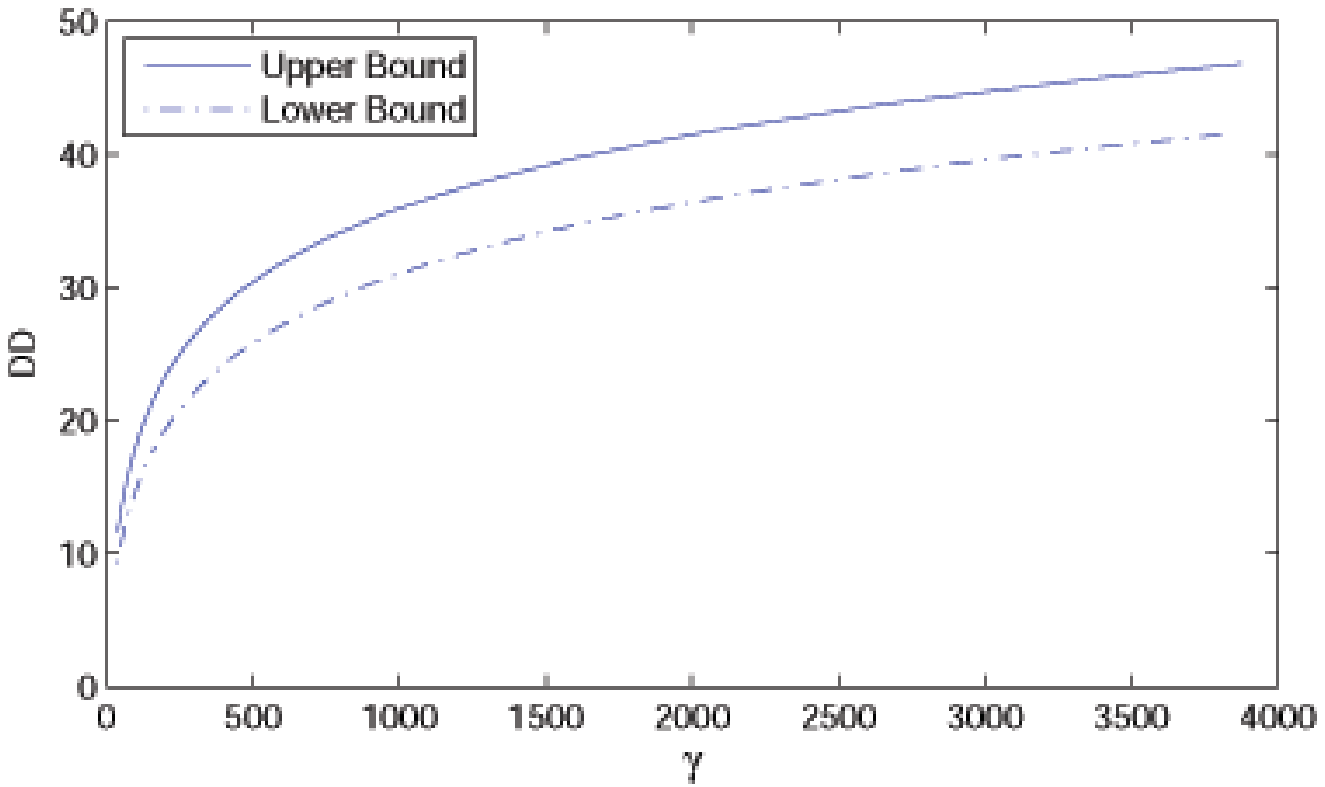}
\label{fig_first_case}} \hfil
\subfloat[$\mu=1$]{\includegraphics[width=2.5in]{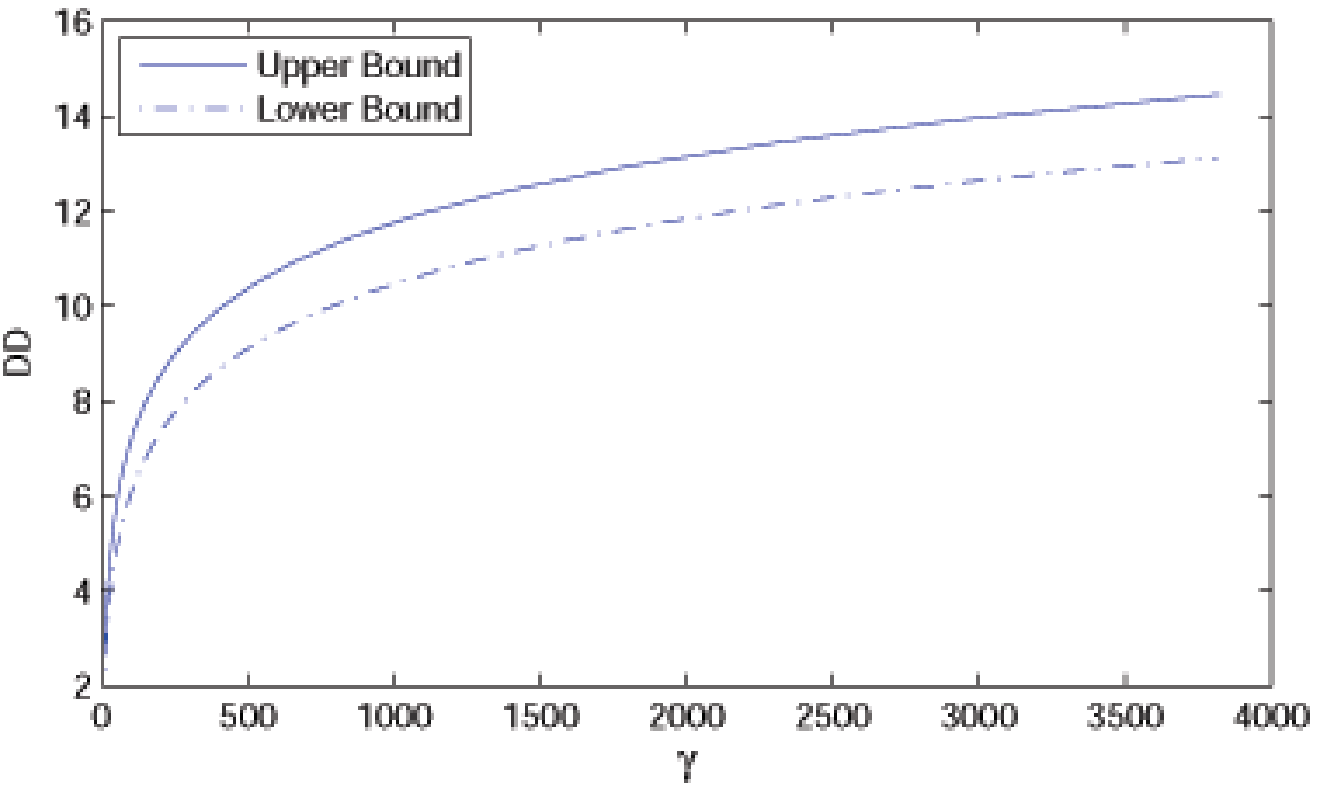} \label{fig_second_case}}
\hfil \subfloat[$\mu=2$]{\includegraphics[width=2.5in]{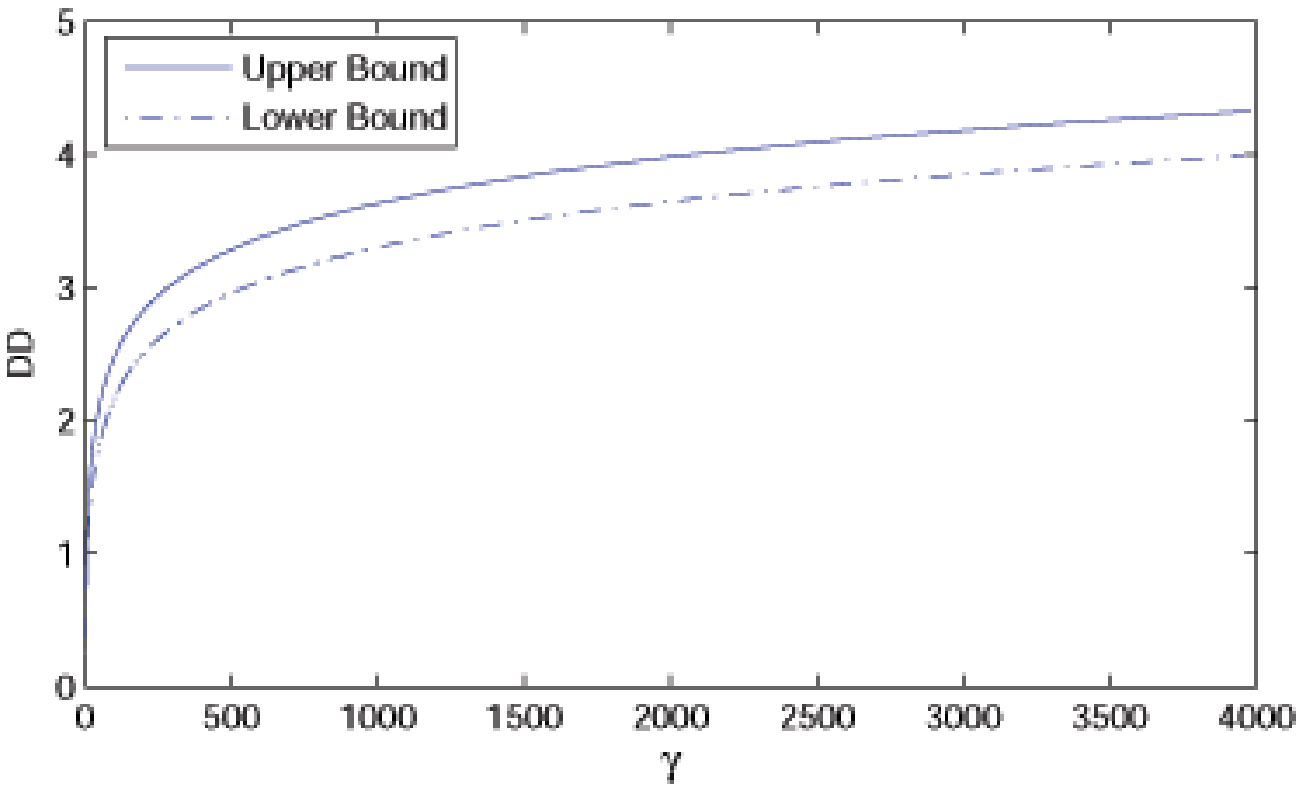}
\label{fig_third_case}} } \caption{{\bf (Left)} Case of $\mu=\frac{1}{2}$. {\bf
(Middle)} Case of $\mu_2=1$. {\bf (Right)} Case of $\mu=2$. (Note
 that the differences between upper and lowers bounds are all bounded as $\gamma$ increases.)} \label{ul}
\end{figure*}

The upper and lower bounds on detection delay for the optimal stopping rule, when
$\mu$ is $\frac{1}{2}$, $1$ and $2$, in the case that $N=2$ are shown in Figure
\ref{ul}.

The consequence of Theorem \ref{main} is the asymptotic optimality of
(\ref{CUSUMmultichart}) in the case in which all of the information becomes directly
available through the filtration $\{\mathcal{F}_t\}$ at the fusion center. We notice
however that this asymptotic optimality holds for any finite number of sensors $N$.

We now 
discuss the implications of the above result in decentralized
detection in the case of one shot schemes.

\section{Decentralized detection}

Let us now suppose that each of the observation processes $\{\xi_t^{(i)}\}$ become
sequentially available at its corresponding sensor $S_i$ which then devises an
asynchronous communication scheme to the central fusion center. In particular,
sensor $S_i$ communicates to the central fusion center only when it wants to signal
an alarm, which is elicited according to a CUSUM rule $T_h^{(i)}$ of
(\ref{CUSUMstop}). Once again the observations received at the $N$ sensors are
independent and can change dynamics at distinct unknown points $\tau_i$. The fusion
center, whose objective is to detect the first time when there is a change, devises
a minimal strategy; that is, it declares that a change has occurred at the first
instance when one of the sensors communicates an alarm. The implication of Theorem
\ref{main} is that in fact this strategy is the best that the fusion center can
devise and that there is no loss in performance between the case in which the fusion
center receives the raw data $\{\xi_t^{(1)},\ldots \xi_t^{(N)}\}$ directly and the
case in which the communication that takes place is limited to the one shown in
Figure \ref{fig1}. To see this, the detection delay of the stopping rule
$T_h=T_h^{(1)} \wedge \ldots \wedge T_h^{(N)}$ is equal to
$E_{0,\infty,\ldots,\infty}\{T_h\}$ when $S_1$ is the one that first signals an
alarm, $E_{\infty,0,\ldots,\infty}\{T_h\}$ when $S_2$ first signals and so on all of
which are equal due to the assumed symmetry in the signal strength $\mu$ received at
each of the sensors $S_i$ when a change occurs. The mean time between false alarms
for the fusion center that devises the rule $T_h=T_h^{(1)} \wedge \ldots \wedge
T_h^{(N)}$ is thus $E_{\infty,\ldots,\infty}\left\{T_h\right\}$. But Theorem
\ref{main} asserts that this rule, namely $T_h$, is asymptotically optimal as the
mean time between false alarms tends to $\infty$ in the centralized case for any
finite $N$. In other words, the CUSUM stopping rules $T_h^{(1)}$, $T_h^{(2)}$, ...,
$T_h^{(N)}$ are sufficient statistics (at least asymptotically) for the problem of
quickest detection of (\ref{eqnproblem}).

\section{Possible extensions}

An interesting extension corresponds to the case in which the signal strengths $\mu$
are different in each sensor after the change. That is, after the change the signal
in $S_i$ is $\mu_i$ with $\mu_1 \ne \mu_2 \ne \ldots \ne \mu_N$. In this case, it is
not clear what the optimal choice of thresholds is, but it is possible that the
thresholds $\{h_i\}$ should be chosen so that
\begin{eqnarray*}
E_{0,\infty,\ldots,\infty}\{T_C^N\}=E_{\infty,0,\infty,\ldots,\infty}\{T_C^N\}=\ldots=E_{\infty,\ldots,\infty,0}\{T_C^N\},
\end{eqnarray*}
where $T_C^N=T_{h_1}^1 \wedge \ldots\wedge T_{h_N}^N$.


A further interesting extension corresponds to the case of correlated sensors. To
demonstrate this case let us begin by assuming that $N=2$. This case corresponds to
(\ref{corr}), but with
\begin{eqnarray}
\label{corrprime} E\left\{w_t^{(1)}w_s^{(2)}\right\} & = & \rho
\min\{s,t\}~~\forall~s,t\ge 0.
\end{eqnarray}
This case becomes significantly more difficult because of the presence of local time
in the dynamics of the process $\max\{y_t^{(1)},y_t^{(2)}\}$. Nevertheless, it is
possible to derive a formula for the expected delay of $T_h$ under the measure
$P_{\infty, \infty}$.  This expression is given by
\begin{eqnarray}\label{E} \nonumber
E_{\infty,\infty}\left\{T_h\right\}& = & \frac{2}{\mu^2}(e^h-h-1) \\ \nonumber  & -
& 2(1-\rho) E\left\{ \int_0^T (e^{y_t^{(1)}}-1) \delta(y_s^{(1)}-y_s^{(2)})
ds\right\}, \\ & ~ &
\end{eqnarray}
where $\delta$ denotes the Dirac delta function and the final term in this
expression corresponds to the collision local time of the processes $y_t^{(1)}$ and
$y_t^{(2)}$ weighted by the factor $(e^{y_t^{(1)}}-1)$. The difficulty in the use of
expression (\ref{E}) is the fact that as $\rho$ changes, the expected value of the
collision local time term, which is the last term in (\ref{E}), also changes.
Moreover, the expression for the first moment of $T_h$ becomes significantly more
complicated under the measure $P_{0,\infty}$.

\newpage

\bibliographystyle{IEEEtran}


\section{Appendix}
As an illustration for general case, let us prove the result for $N=2$.

We begin by deriving expressions for $E_{0,\infty}\left\{T_h
\right\}$ and $E_{\infty,\infty}\left\{T_h \right\}$ by using the
results in \cite{Malik}. For all $h> 2$, we have
\begin{multline*}
E_{0,\infty}\left\{T_h
\right\} =\\
 \frac{32}{\mu^2}\sum_{i,j\geq
1}\frac{\sin^3\phi_i\sin^3\theta_j}{(\phi_i-\sin\phi_i\cos\phi_i)(\theta_j-\sin\theta_j\cos\theta_j)}\frac{\cos^2\phi_i\cos^2\theta_j}{\cos^2\phi_i+\cos^2\theta_j} \\
-\frac{32}{\mu^2}\frac{\sinh^3\eta}{\sinh\eta\cosh\eta-\eta}\sum_{i\geq1}\frac{\sin^3\phi_i\cos^2\phi_i}{\phi_i-\sin\phi_i\cos\phi_i}\frac{ \cos^2\phi_i}{\cos^2\phi_i+\cosh^2\eta} \\
+\frac{32}{\mu^2}\frac{\sinh^3\eta}{\sinh\eta\cosh\eta-\eta}\sum_{i\geq1}\frac{\sin^3\phi_i\cos^2\phi_i}{\phi_i-\sin\phi_i\cos\phi_i}\\
 = S_1(h)+S_2(h)+S_3(h),
\end{multline*}
and
\begin{multline*}
E_{\infty,\infty}\left\{T_h \right\}=\\
\frac{32}{\mu^2}\sum_{i,j\geq
1}\frac{e^{-h}\sin^3\theta_i\sin^3\theta_j}{(\theta_i-\sin\theta_i\cos\theta_i)(\theta_j-\sin\theta_j\cos\theta_j)}\frac{\cos^2\theta_i\cos^2\theta_j}{\cos^2\theta_i+\cos^2\theta_j}\\
+\frac{64}{\mu^2}\frac{e^{-h}\sinh^3\eta}{\sinh\eta\cosh\eta-\eta}\sum_{i\geq
1}\frac{\sin^3\theta_i\cos^2\theta_i}{\theta_i-\sin\theta_i\cos\theta_i}\frac{\cosh^2\eta}{\cos^2\theta_i+\cosh^2\eta}\\
+\frac{16}{\mu^2}\frac{e^{-h}\sinh^6\eta\cosh^2\eta}{(\sinh\eta\cosh\eta-\eta)^2}\\
=S_4(h)+S_{5}(h)+S_{6}(h),
\end{multline*}
where
\begin{eqnarray*}
\tan\phi_n&=&-\frac{2}{h}\phi_n<0 \\
\tan\theta_n &=& \frac{2}{h} \theta_n>0\\
\tanh \eta&=&\frac{2}{h} \eta>0.
\end{eqnarray*}

The idea then is show $S_1(h)$, $S_2(h)$, $S_4(h)$ and $S_5(h)$
converge to zero, and examine how $S_2(h)$ and $S_5(h)$ behave as $h
\to \infty$. In the following paragraphs we shall analyze these in
the order $S_6(h)\rightarrow S_3(h)\rightarrow
S_2(h)\rightarrow S_5(h)\rightarrow S_1(h)\rightarrow S_4(h)$.\\

First notice that for large $h$, $\eta$ is large and close to
$\frac{h}{2}$. Moreover,
\begin{eqnarray}
\label{hneta}e^{2\eta-h}=1-4\eta e^{-2\eta}+o(e^{-3\eta}).
\end{eqnarray}
This can help us to compare $\eta$ with $h$.

For $S_6(h)$,
\begin{eqnarray*}
S_6(h)&=&\frac{16}{\mu^2}e^{-h}\sinh^4\eta\left(1-\eta\sinh^{-1}\eta\cosh^{-1}\eta\right)^{-2}\\
&=&\frac{1}{\mu^2}e^{-h}\left[e^{4\eta}+\left(8\eta-4\right)e^{2\eta}+o(e^{\eta})\right]\\
&=&\frac{1}{\mu^2}\left[e^he^{4\eta-2h}+(8\eta-4)e^{2\eta-h}+o(e^{-\eta})\right],
\end{eqnarray*}
by (\ref{hneta}) the first term is
\begin{eqnarray*}
e^he^{4\eta-2h}&=&e^h\left(e^{2\eta-h}\right)^2=e^h\left[1-4\eta
e^{-2\eta}+o(e^{-3\eta})\right]^2\\
&=&e^h\left[1-8\eta e^{-2\eta}+o(e^{-3\eta})\right]\\&=&e^h-8\eta
e^{h-2\eta}+o(e^{-\eta})\\
&=&e^h-8\eta+o(e^{-\eta}),
\end{eqnarray*}
and the second term is
\begin{eqnarray*}
(8\eta-4)e^{2\eta-h}&=&(8\eta-4)\left[1+o(e^{-\eta})\right]\\&=&8\eta-4+o(e^{-\eta}),
\end{eqnarray*}
so
\begin{eqnarray}
\label{s6}S_6(h) & = &
\frac{1}{\mu^2}\left[e^h-4+o(e^{-\frac{h}{2}})\right],~\textrm{as}~h \to \infty.
\end{eqnarray}

For $S_3(h)$, also note that from (\ref{DD}) and \cite{Malik} we can write
\begin{eqnarray}
\nonumber \frac{2}{\mu^2}f(-h)&=&\int_{0}^{\infty}P_0(T_h>t)dt\\
&=&\frac{16}{\mu^2}e^{\frac{h}{2}}\sum_{i\ge1}\frac{\sin^3\phi_i\cos^2\phi_i}{\phi_i-\sin\phi_i\cos\phi_i},
\end{eqnarray}
from which we obtain
\begin{eqnarray}\label{s3}
\nonumber S_3(h)&=&\frac{32}{\mu^2}\frac{\sinh^3\eta}{\sinh\eta\cosh\eta-\eta}\sum_{i\ge1}\frac{\sin^3\phi_i\cos^2\phi_i}{\phi_i-\sin\phi_i\cos\phi_i}\\
\nonumber&=&\frac{2}{\mu^2}\left[1+o(e^{-\eta})\right](h+e^{-h}-1)\\
&=&\frac{2}{\mu^2}\left[h-1+o(e^{-\frac{h}{2}})\right],~\textrm{as}~h \to \infty.
\end{eqnarray}

To bound $S_2(h)$ and $S_5(h)$ we need the following,

\textit{Result 1: }Suppose $0<p<1$. Then, for all positive solutions
$\{\alpha_i\}_{i\ge1}$ to the equation $\tan x=px$ ($\tan x=-px$, \textit{resp.}),
we have
\begin{eqnarray}
\label{result1}\lim_{p\to0^{+}}\sum_{i\ge1}\frac{|\sin^3\alpha_i|\cos^2\alpha_i}{\alpha_i-\sin\alpha_i\cos\alpha_i}\leq\frac{1}{\pi}.
\end{eqnarray}

This suggests that, asymptotically, as $h \to \infty$,
\begin{multline*}
\frac{\sinh^3\eta}{\sinh\eta\cosh\eta-\eta}\sum_{i\ge1}\frac{|\sin^3\phi_i|\cos^2\phi_i}{\phi_i-\sin\phi_i\cos\phi_i}\frac{\cos^2\phi_i}{\cos^2\phi_i+\cosh^2\eta}\\
\le
\left[\frac{1}{\pi}+o(1)\right]\frac{\sinh^3\eta}{\cosh^2\eta(\sinh\eta\cosh\eta-\eta)}\\
=\left[\frac{1}{\pi}+o(1)\right]\frac{\sinh^2\eta}{\cosh^3\eta}\left(1-\eta\sinh^{-1}\eta\cosh^{-1}\eta\right)^{-1}\\
=O(e^{-\frac{h}{2}}),
\end{multline*}
from which we obtain
\begin{eqnarray}
\label{s2}|S_2(h)| & = & \frac{2}{\mu^2}O(e^{-\frac{h}{2}}),~\textrm{as}~h \to
\infty.
\end{eqnarray}

Similarly,
\begin{multline*}
\frac{e^{-h}\sinh^3\eta}{\sinh\eta\cosh\eta-\eta}\sum_{i\ge1}\frac{|\sin^3\theta_i|\cos^2\theta_i}{\theta_i-\sin\theta_i\cos\theta_i}\frac{\cosh^2\eta}{\cos^2\theta_i+\cosh^2\eta}\\
\le\left[\frac{1}{\pi}+o(1)\right]e^{-h}\frac{\sinh^2\eta}{\cosh\eta}\left(1-\eta\sinh\eta\cosh\eta\right)^{-1}\\
=O(e^{-\frac{h}{2}}),
\end{multline*}
so
\begin{eqnarray}
\label{s5}|S_5(h)| & = &
\frac{1}{\mu^2}O(e^{-\frac{h}{2}}),~\textrm{as}~h \to \infty.
\end{eqnarray}

To handle the double sum in $S_1(h)$ and $S_4(h)$, we need

\textit{Result 2: }Suppose $0<p<1$, $\{\alpha_i\}_{i\ge1}$ are all
positive solutions to the equation $\tan x=px$, and $\{\beta_i\}_{i
\ge 1}$ are all positive solutions to equation $\tan x=px$ ($\tan
x=-px$, \textit{resp.}), then
\begin{eqnarray}\label{result2}
\nonumber\sum_{i,j\ge1}\frac{\sin^3\alpha_i\sin^3\beta_j}{(\alpha_i-\sin\alpha_i\cos\alpha_i)(\beta_j-\sin\beta_j\cos\beta_j)}
\frac{\cos^2\alpha_i\cos^2\beta_j}{\cos^2\alpha_i+\cos^2\beta_j}\\
\to0,~\textrm{as}~p\to 0^+.
\end{eqnarray}
Consequently,
\begin{eqnarray}
\label{s1}|S_1(h)|&=&o(1),~\textrm{as}~h \to \infty.
\end{eqnarray}
Similarly,
\begin{eqnarray}
\label{s4}|S_4(h)| & = & \frac{1}{\mu^2}o(e^{-h}),~\textrm{as}~h \to
\infty.
\end{eqnarray}

Finally, from (\ref{s6}), (\ref{s3}), (\ref{s2}), (\ref{s5}),
(\ref{s1}) and (\ref{s4}) we obtain
\begin{eqnarray}
\label{E0inf}
\hspace*{3ex}E_{0,\infty}(T_h)&=&\frac{2}{\mu^2}\left[h-1+o(1)\right],
~\textrm{as}~h \to \infty,\end{eqnarray} and
\begin{eqnarray}
\label{Einfinf}E_{\infty,\infty}(T_h)&=&\frac{1}{\mu^2}\left[e^h-4+o(1)\right],
~\textrm{as}~h \to \infty.
\end{eqnarray}
And for $h$ and $\gamma$ satisfying (\ref{FACSh}), we have
asymptotic results (\ref{AsymptoticSh}) with $N=2$.

Now let us prove the two results we used in the above.\\
\noindent Result 1:\\
\begin{proof}
For any
$\alpha_i\in\left((i-\frac{1}{2})\pi,(i+\frac{1}{2})\pi\right)$ such
that $\tan\alpha_i=\pm p\alpha_i$, ($0<p<1$), we have
\begin{multline*}
\nonumber\frac{|\sin^3\alpha_i|\cos^2\alpha_i}{\alpha_i-\sin\alpha_i\cos\alpha_i}=\frac{p^3\alpha_i^2}{(1+p^2\alpha_i^2)^{3/2}\left[(1\mp
p)+p^2\alpha_i^2\right]}\\
\leq\frac{p}{(1+p^2\alpha_i^2)^{3/2}}
\leq\frac{p}{\left[1+p^2\left(i-\frac{1}{2}\right)^2\pi^2\right]^{3/2}}.
\end{multline*}
Thus
\begin{multline*}
\sum_{i\ge1}\frac{|\sin^3\alpha_i|\cos^2\alpha_i}{\alpha_i-\sin\alpha_i\cos\alpha_i}\leq\sum_{i\geq1}\frac{p}{\left[1+p^2\left(i-\frac{1}{2}\right)^2\pi^2\right]^{3/2}}\\
\leq\frac{1}{\pi}\int_{-\frac{\pi}{2}}^{\infty}\frac{pdx}{(1+p^2x^2)^{3/2}}
=\frac{1}{\pi}\int_{-p\frac{\pi}{2}}^{\infty}\frac{du}{(1+u^2)^{3/2}}\\
\to\int_{0}^{\infty}\frac{du}{(1+u^2)^{3/2}}=\frac{1}{\pi},~\textrm{as}~p\to0^{+}.
\end{multline*}
\end{proof}

\noindent Result 2:\\
\begin{proof}
For simplicity let us denote the $(i,j)$-term in the sum by $a_{i,j}(p)$. As in the
last proof, a little computation would give us
\begin{eqnarray*}
|a_{i,j}(p)|&=&I_p(p\alpha_i,p\beta_j)\cdot p^2,
\end{eqnarray*}
where $I_p(x,y)$ $(0<p<1)$ is the function
\begin{multline*}
I_p(x,y)=\\
\frac{1}{\sqrt{(1+x^2)(1+y^2)}(2+x^2+y^2)(1+\frac{1-p}{x^2})(1+\frac{1\mp
p}{y^2})}.
\end{multline*}
Clearly, $I_p(\cdot,\cdot)$ is (uniformly in $p$, $0\le p<1$)
bounded above by the $L^1(\mathbb{R}^2)$ function $B(\cdot,\cdot)$,
which is defined as
\begin{eqnarray*}
B(x,y)&=&\frac{1}{\sqrt{(1+x^2)(1+y^2)}(2+x^2+y^2)}.
\end{eqnarray*}

 We have two steps to finish our proof:\\
\begin{multline*}
(a)
\lim_{p\to0^+}\sum_{i,j\ge1}|a_{i,j}(p)|=\frac{1}{\pi^2}\int\int_{(\mathbb{R}^+)^2}I_0(x,y)dxdy,
\end{multline*}
\begin{multline*}
(b)
\lim_{p\to0^{+}}\sum_{i,j\ge1}a_{i,j}(p)\cdot\mathbb{I}_{\{a_{i,j}(p)>0\}}\\=\frac{1}{2\pi^2}\int\int_{(\mathbb{R}^+)^2}I_0(x,y)dxdy.
\end{multline*}

 Let us start from (a). Given any $\epsilon>0$, we can find a constant
$M>0$ such that, for $R_M=\{(x,y):\min(x,y)>M\}$ and all $0\le p<1$,
\begin{displaymath}
\begin{cases}
& I_p~\textrm{is decreasing in
both}~x~\textrm{and}~y~\textrm{in}~R_M,\\
&\frac{1}{\pi^2}\int\int_{R_M}I_p(x,y)dxdy\le\frac{1}{\pi^2}\int\int_{R_M}B(x,y)dxdy\le\frac{\epsilon}{3}.
\end{cases}
\end{displaymath}
Because of this, for any $0\le p<1$, the ``tail'' sum
\begin{eqnarray}\label{tail}
\nonumber\sum_{\min(p\alpha_i,p\beta_j)>M+p\pi}|a_{i,j}(p)|\le\frac{1}{\pi^2}\int\int_{R_M}I_p(x,y)dxdy\le\frac{\epsilon}{3},\\
\end{eqnarray}
where we define $a_{i,j}(0)=0$ for all $(i,j)$.

\begin{figure*}
\normalsize \setcounter{MYtempeqncnt}{\value{equation}} \setcounter{equation}{37}
\begin{equation}\label{ncensors}
\sum_{i,j\ge1}\frac{\sin^3\phi_i\sin^3\theta_{j}\cos^2\phi_i\cos^2\theta_j}{(\phi_i-\sin\phi_i\cos\phi_i)(\theta_{j}-\sin\theta_{j}\cos\theta_{j})[(N-2)(1-4\frac{\eta^2}{h^2})\cos^2\phi_i\cos^2\theta_j+\cos^2\phi_i+\cos^2\theta_j]}\\
\end{equation}
\begin{equation}\label{ncenfun}
I_p^{(N)}(x,y)=\frac{1}{\sqrt{(1+x^2)(1+y^2)}[(N-2)(1-p^2\eta^2)+2+x^2+y^2](1+\frac{1-p}{x^2})(1+\frac{1+p}{y^2})}
\end{equation}
\setcounter{equation}{\value{MYtempeqncnt}} \hrulefill \vspace*{4pt}
\end{figure*}

 On the other hand, as $p$ goes to zero, the function $I_p$
will converge uniformly in $[0,M]^2$ to $I_0$. So all the terms
$|a_{i,j}(p)|$ in the ``head'' sum are uniformly very close to
$I_0(p\alpha_i,p\beta_j)\cdot p^2$, the sum of which, multiplied by
$\pi^2$, is a Riemann sum of the function $I_0(x,y)$ over the region
$[0,M]^2$, and will converges to the Riemann integral of $I_0$ over
$[0,M]^2$ as $p$ turns to zero. In other words, for small $p$, there
exists
\begin{eqnarray}\label{head}
\nonumber\left|\sum_{\max(p\alpha_i,p\beta_j)\le
M+p\pi}|a_{i,j}(p)|-\frac{1}{\pi^2}\int_0^M\int_{0}^{M}I_0(x,y)dxdy\right|\\
\le\frac{\epsilon}{3}.
\end{eqnarray}
By (\ref{tail}) and (\ref{head}), we have
\begin{eqnarray}
\nonumber\left|\lim_{p\to0^+}\sum_{i,j\ge1}|a_{i,j}(p)|-\frac{1}{\pi^2}\int\int_{(\mathbb{R}^+)^2}I_0(x,y)dxdy\right|
\le\epsilon.\\
\end{eqnarray}
Now let $\epsilon$ goes to zero we are done with (a).

 The proof of (b) is similar. Note that the signs of the $a_{i,j}(p)$'s
can be represented by $(-1)^{i+j}$ or $(-1)^{i+j+1}$, and in each
rectangle $[2(i-1)p\pi,2ip\pi]\times[(j-1)p\pi,jp\pi], (i,j\ge1)$,
either $a_{2i-1,j}(p)$ or $a_{2i,j}(p)$ is positive. With the same
constant $M$ chosen as above, for the sum of all positive
$a_{i,j}(p)$'s such that $\max(p\alpha_i,p\beta_j)\le M+p\pi$, we
can use the same argument as before, to show that for small $p$,
\begin{eqnarray}\label{posisum}
\nonumber2\pi^2\left(\sum_{\max(p\alpha_i,p\beta_j)\le
M+p\pi}a_{i,j}(p)\cdot\mathbb{I}_{\{a_{i,j}(p)>0\}}\right)\\
\approx\int_0^M\int_0^M I_0(x,y)dxdy.
\end{eqnarray}
Thus (b) is proven because both the tail integral and the tail sum
are negligible due the way to choose $M$.
\end{proof}

In the $N$ CUSUMs case with $N\ge2$, the calculation is similar:
both of the main terms in $E_{0,\infty,\ldots,\infty}\left\{T_h
\right\}$ and in $E_{\infty,\infty,\ldots,\infty}\left\{T_h
\right\}$ are the terms with highest degree in
$\left[\sinh^3\eta/(\sinh\eta\cosh\eta-\eta)\right]$. With
(\ref{hneta}) we can get they are
\begin{eqnarray}\label{main0}
\frac{2}{\mu^2}[h-1+o(e^{-\frac{h}{2}})],
\end{eqnarray}
and
\begin{eqnarray}\label{maininf}
\frac{2}{N\mu^2}\left[e^h+(N-2)h+(2-3N)+o(e^{-\frac{h}{2}})\right],
\end{eqnarray}
respectively.

We can prove that all other terms converge to zero as $h$ goes to
infinity.

With a generalization of {Result 1} to $n$ dimensional trigonometric sums and
integrals for all $n\ge1$, we are able to deal with most terms in the expansion of
the expectations, because those bounded trigonometric sums are multiplied by
expressions of negative exponential order in $h$.

There is only one term (in $E_{0,\infty,\ldots,\infty}\left\{T_h \right\}$) which
cannot be proven to converge to zero in this manner. We need to prove the sum
involved there, which is (\ref{ncensors}) at the top of the following page,
converges to zero as $h$ goes to infinity. We can follow the proof of Result 2 to
get the result. To be more precise, denote $p=\frac{2}{h}$, and the term in above
sum by $a_{i,j}^{(N)}(p)$, then obviously,
$|a_{i,j}^{(N)}(p)|\le|a_{i,j}^{(2)}(p)|$, that can help us to control the ``tail''
sum
\begin{eqnarray}
\setcounter{equation}{40}
\sum_{\min(p\phi_i,p\theta_j)>M+p\pi}|a_{i,j}^{(N)}(p)|,
\end{eqnarray}
where $M$ is chosen as in the proof of Result 2. On the other hand,
\begin{eqnarray}
|a_{i,j}^{(N)}(p)|=I_p^{(N)}(p\phi_i,p\theta_j)\cdot p^2,
\end{eqnarray}
where $I_p^{(N)}$ is the function defined in (\ref{ncenfun}). The
function $I_p^{(N)}$ uniformly converges to $I_0$ as $p$ goes to
infinity in the domain $[0,M]^2$, since $p\eta\to1$ as $p\to0^+$. As
a result, the ``head'' sum converges to the same double integral as
the one in (\ref{head}) or (\ref{posisum}), so we are done!

Finally, by (\ref{main0}) and (\ref{maininf}), we can derive
asymptotic formula (\ref{AsymptoticSh}) with $h$ and $\gamma$
satisfying (\ref{FACSh}).

\end{document}